\documentclass[sigconf, nonacm]{acmart}
\setkeys{acmart.cls}{pbalance=true}
\usepackage{booktabs}%
\usepackage{multirow}
\usepackage{enumitem}
\usepackage{subcaption}%
\usepackage{algorithm}
\usepackage{algpseudocode}%
\usepackage{xcolor}
\usepackage{colortbl}
\usepackage{tabularx}
\definecolor{oursrowblue}{HTML}{EAF2FA}
\usepackage{makecell}
\usepackage{placeins}%
\usepackage{pgfplots}
\pgfplotsset{compat=1.18}
\usepgfplotslibrary{groupplots}

\newsavebox{\shrinkbox}
\newcommand{\shrinktofit}[2]{%
  \sbox{\shrinkbox}{#2}%
  \ifdim\wd\shrinkbox>#1\relax\resizebox{#1}{!}{\usebox{\shrinkbox}}%
  \else\usebox{\shrinkbox}\fi}
\title{Decision-Native Recommendation with Large Language Models}
\author{Daoxuan Liu}
\authornote{Equal contribution.}
\email{daoxuanliu@mail.ustc.edu.cn}
\affiliation{\institution{University of Science and Technology of China}\city{Hefei}\country{China}}
\author{Jiayi Liao}
\authornotemark[1]
\email{ljy0ustc@mail.ustc.edu.cn}
\affiliation{\institution{Independent Researcher}\city{Beijing}\country{China}}
\author{Jiancan Wu}
\email{wujcan@gmail.com}
\affiliation{\institution{University of Science and Technology of China}\city{Hefei}\country{China}}
\author{An Zhang}
\email{an.zhang3.14@gmail.com}
\affiliation{\institution{University of Science and Technology of China}\city{Hefei}\country{China}}
\author{Xiang Wang}
\email{xiangwang1223@gmail.com}
\affiliation{\institution{University of Science and Technology of China}\city{Hefei}\country{China}}
\author{Xiangnan He}
\email{xiangnanhe@gmail.com}
\affiliation{\institution{University of Science and Technology of China}\city{Hefei}\country{China}}
\renewcommand{\shortauthors}{Liu et al.}
\setcopyright{none}
\acmDOI{}
\acmISBN{}
\acmConference{}{}{}
\acmBooktitle{}
\hypersetup{pdfauthor={Daoxuan Liu, Jiayi Liao, Jiancan Wu, An Zhang, Xiang Wang, Xiangnan He}, pdfkeywords={sequential recommendation, large language models, confidence calibration}}
\begin{document}
\begin{abstract}
Large language models (LLMs) bring world knowledge to user preference modeling in sequential recommendation, but generation-based recommenders express next-item predictions indirectly as text.
Generated outputs may fail to identify a unique candidate item, depend on candidate presentation order in the prompt, and lack explicit confidence in the recommendation.
Recent studies explore Jev, a model for fast option selection with explicit confidence.
Inspired by Jev, we introduce decision-native recommendation, a new paradigm that jointly models item choice and recommendation confidence using LLM representations.
It requires valid candidate selection, consistency under candidate reordering, and confidence that reflects recommendation accuracy.
We propose Direct Item Choice Estimation (DICE), which learns a personalized next-item distribution over supplied candidates from user interaction history and recommends the highest-scoring item with calibrated confidence.
Tree-structured attention shares history computation across isolated candidate branches for order-consistent candidate scoring in a single forward pass.
On three recommendation datasets, DICE achieves higher top-one accuracy than all evaluated baselines.
Separate evaluations show unchanged item choices under tested candidate reorderings and end-to-end speedups of up to approximately $6\times$ over the evaluated generation-based methods.
Its confidence estimates support selective recommendation by identifying more reliable predictions and guide additional computation for low-confidence examples in offline routing.
\end{abstract}
\maketitle
\hypersetup{pdfauthor={Daoxuan Liu, Jiayi Liao, Jiancan Wu, An Zhang, Xiang Wang, Xiangnan He}, pdfkeywords={sequential recommendation, large language models, confidence calibration}}
\section{Introduction}
\label{sec:introduction}
Recommendation on Web platforms is a decision about which items to present to a
user. This decision has conventionally been learned as classification over items.
Sequential recommenders map a user's interaction history to a distribution over items~\cite{hidasi2016gru4rec,tang2018caser,kang2018sasrec}, and click-through-rate (CTR) models estimate the probability that a user engages with a given item~\cite{guo2017deepfm,wang2017dcn,zhou2018din}.
These recommenders score items directly using representations learned from interaction data and task-specific features.
Large language models (LLMs) expand the knowledge available for recommendation by interpreting item descriptions and user histories through representations learned during pretraining~\cite{geng2022p5,bao2023tallrec,liao2023llara}.
To exploit this knowledge, most LLM-based recommenders adopt a generation-based paradigm, autoregressively generating the next item's name or a ranked list of candidate names from a prompt~\cite{bao2023bigrec,chen2024sdpo,ding2025napo}.
\begin{figure}[t]
  \centering
  \includegraphics[width=\columnwidth]{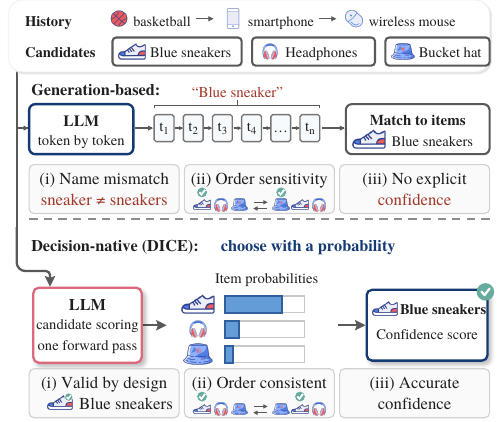}
  \caption{Generation-based recommendation and DICE for the same history and candidates. Generated item names require candidate matching. DICE selects a supplied candidate and returns calibrated confidence in that choice. Probability bars are illustrative.}
  \label{fig:choice-interface}
  \Description{}

\end{figure}
This paradigm gains world knowledge but gives up the decision form of recommendation, in which the output is a choice among items.
Figure~\ref{fig:choice-interface} illustrates three concerns with generation-based recommendation. First, the generated text has to be mapped back to the item space~\cite{chen2024sdpo,bao2023bigrec,lin2023transrec}, and in the experiment, roughly 8--16\% of generated answers do not identify a unique candidate item (Section~\ref{sec:rq2}).
Second, the output is sensitive to how the candidates are presented~\cite{hou2023llmrank,bito2025positionbias,bito2026preferenceconsistency}.
With the user's interaction history and candidate set held fixed, merely reordering the candidate items in the prompt changes the selected item or answer validity in roughly 30--45\% of tested reorderings for the evaluated generation-based baselines (Table~\ref{tab:permutation}).
Third, the generated item name alone does not provide explicit confidence in how likely the recommendation is to be correct.
In addition, token-by-token decoding adds considerable inference latency.
Yet recent work uses Jev~\cite{almeida2026jev}, a System-1 model designed to make fast decisions with explicit confidence rather than generate text~\cite{wu2026reflex,li2026jevasjudge,zhu2026jevagentharness}.
Given a context and a set of options, Jev directly selects an option and returns a confidence score for that choice.
Inspired by this approach, we argue that using LLM knowledge for recommendation does not require language generation.
We therefore introduce \textbf{decision-native recommendation}, a new paradigm that uses LLM representations to jointly model item choice and confidence.
Under this paradigm, a recommender should select a valid candidate, preserve its choice under candidate reordering, and provide confidence estimates that accurately reflect how likely that choice is to be correct.
Realizing this paradigm raises two challenges.
First, the model must learn which supplied item a user is likely to interact with next and estimate confidence in that choice. This requires learning preferences over candidate items from the user's interaction history, while accurate next-item prediction alone does not ensure reliable confidence.
Second, the architecture must preserve the temporal order of past interactions without making candidate scores depend on presentation order. A straightforward way to score each candidate is to run the model separately on the user's interaction history and that candidate. However, this processes the same history repeatedly, once for every candidate. The challenge is to share this computation while preserving the history's chronology and keeping candidate scores independent of presentation order.
To address these challenges, we propose \textbf{Direct Item Choice Estimation (DICE)} for personalized next-item recommendation.
A pretrained LLM encodes each candidate's description together with the user's chronological interaction history, and a shared scoring head estimates the user's preference for that candidate.
These scores are normalized into a choice distribution, trained with candidate-level cross-entropy to favor the observed next item over the alternatives.
At inference, DICE recommends the highest-scoring candidate.
To estimate confidence in this choice, temperature scaling fitted on held-out data calibrates the choice probabilities without changing the selected item.
The recommended item's calibrated probability then estimates how likely that recommendation is to be correct.
Tree-structured attention computes all candidate scores in one forward pass by sharing history computation while blocking attention between candidates.
The history retains its temporal order, and consistent position assignments keep candidate scores independent of candidate presentation order.
We evaluate DICE on three public recommendation datasets.
First, DICE achieves the best top-one recommendation accuracy among the evaluated baselines.
Additional experiments show accuracy gains with larger backbones and support selection of items unseen during recommendation training (Section~\ref{sec:results}).
Second, DICE returns valid candidate items, preserves its choices under all tested candidate reorderings, and achieves an end-to-end speedup of up to approximately $6\times$ over the evaluated generation-based methods (Section~\ref{sec:rq2}).
Third, calibrated decision confidence helps identify recommendations that are more likely to be correct.
On two benchmarks, DICE's most confident recommendations, covering roughly a quarter of each test set, achieve 88.82--94.42\% accuracy.
In an offline routing evaluation, escalating fewer than half of the examples to a 4B model retains over 99.5\% of that model's accuracy with 24.8\% lower estimated sequential latency (Section~\ref{sec:rq3}).
We summarize our contributions as follows:
\begin{itemize}[leftmargin=*]
  \item We introduce \emph{decision-native recommendation}, a new paradigm for LLM-based recommendation that retains the world knowledge of LLMs while recovering the decision form of recommendation, and we characterize the challenges in realizing it.
  \item We propose \textbf{DICE}, which instantiates the paradigm with user--item encoding, order-consistent categorical choice, single-pass scoring over a shared user history, and calibrated decision confidence.
  \item We show empirically that DICE outperforms all evaluated baselines in top-one accuracy across three benchmarks. Further evaluations demonstrate efficient inference and the value of decision confidence for selective recommendation and model routing.
\end{itemize}

\section{Related Work}
\label{sec:related}
\subsection{LLM-Based Recommendation}
LLM-based recommenders model preferences using interaction histories, item text, and pretrained knowledge~\cite{wu2023llm4recsurvey,li2024genrecsurvey,wang2024nextgenllmrec}.
Representation methods include UniSRec and Recformer~\cite{hou2022unisrec,li2023recformer}; LLaRA, CoLLM, and A-LLMRec further integrate sequential or collaborative embeddings with LLM inputs~\cite{liao2023llara,zhang2023collm,kim2024allmrec}.
Generation-based methods include P5's text-to-text framework~\cite{geng2022p5}, GenRec~\cite{ji2023genrec}, recommendation tuning in TALLRec~\cite{bao2023tallrec}, zero-shot ranking in LLMRank~\cite{hou2023llmrank}, and preference supervision in S-DPO and NAPO~\cite{chen2024sdpo,ding2025napo}.
BIGRec and TransRec address grounding generated outputs in the item space~\cite{bao2023bigrec,lin2023transrec}.
Reasoning-enhanced methods add rationales or reasoning graphs~\cite{tsai2024reasoningrec,wang2023reasoninggraphs} or optimize reasoning--answer trajectories~\cite{re2code}.
RDRec distills rationales~\cite{wang2024rdrec}, SelfDR learns from a rationale-informed teacher without test-time reasoning~\cite{jiang2026selfdr}, and Think When Needed invokes reasoning selectively~\cite{guo2026thinkwhenneeded}.
DICE implements decision-native recommendation by learning candidate-choice probabilities from contextual representations and evaluating the selected item's confidence together with output validity and order consistency.
\subsection{Direct Scoring and Efficient Inference}
Direct LLM scoring avoids answer generation.
RankLLaMA scores query--document relevance from contextual representations~\cite{ma2024rankllama}, while LlamaRec maps candidate-letter logits to item scores~\cite{yue2023llamarec}.
InvariRank combines candidate isolation, shared positional framing, and mean token-log-probability scoring for order-consistent ranking in a single forward pass~\cite{bito2026invarirank}.
We introduce decision-native recommendation as a new paradigm that treats valid candidate selection, candidate-order consistency, and confidence that reflects observed recommendation accuracy as joint requirements.
DICE instantiates this paradigm by learning candidate-choice probabilities from LLM representations and calibrating confidence in the selected item.
We compare its ranking performance with LlamaRec and InvariRank and assess all three methods' calibration before and after method-specific held-out temperature scaling.
SpecInfer and Medusa use tree-structured attention to verify proposed token continuations in parallel~\cite{miao2024specinfer,cai2024medusa}.
DICE-I adapts this computation pattern to share history computation across isolated candidate branches while preserving candidate-order consistency.
\subsection{Calibration, Selective Prediction, and Routing}
Temperature scaling calibrates neural classifiers and pretrained Transformers~\cite{guo2017calibration,desai2020calibration}.
LLM confidence studies compare conditional probabilities with verbalized confidence~\cite{tian2023calibration} and evaluate confidence elicitation~\cite{xiong2024uncertainty}.
Confidence also supports selective prediction, which studies error at different coverage levels~\cite{geifman2017selective,kamath2020selectiveqa}.
For computation allocation, FrugalGPT learns model cascades~\cite{chen2023frugalgpt}, RouteLLM routes using preference data~\cite{ong2025routellm}, and language-model cascades defer using token-level uncertainty~\cite{gupta2024cascades}.
At the level of individual decisions, System-1.x allocates search by subproblem difficulty~\cite{saha2024systemonex}, while REFLEX, Jev-Mem, and JEV-as-a-Judge apply System-1 control to agent actions, memory operations, and selective judgments~\cite{wu2026reflex,jiang2026jevmem,li2026jevasjudge}.
DICE applies established temperature scaling to its candidate-choice distribution and evaluates how well selected-item confidence reflects observed recommendation accuracy.

\section{Method}
\label{sec:method}
\begin{figure*}[t]
  \centering
  \includegraphics[width=\textwidth]{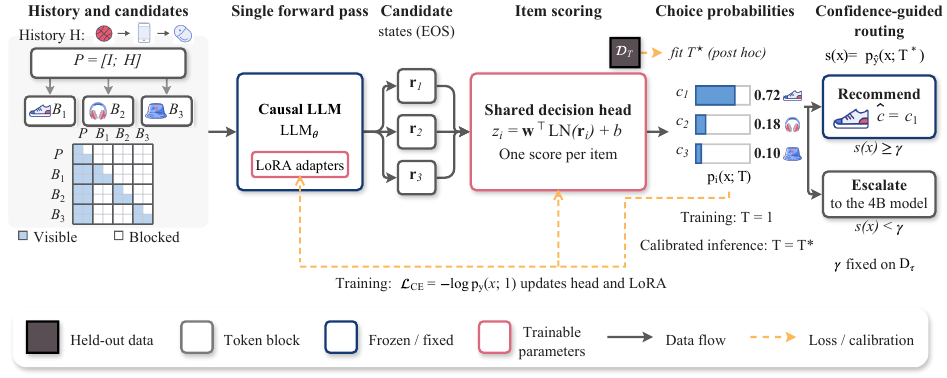}
  \caption{DICE-I scoring and confidence-guided routing. Icons represent item texts. Mask rows attend to columns, with causal attention within $P$ and each $B_i$. Branches read $P$ but not one another. Branch positions match independent encoding. LoRA and the head train at $T=1$. Calibration fits $T^*$ on $D_T$ with model parameters frozen. The routing threshold $\gamma$ is fixed on disjoint $D_\tau$. Token counts and probabilities are illustrative.}
  \Description{The left-to-right pipeline shows history and candidate inputs, a single causal LLM forward pass, candidate states at EOS, item scoring, choice probabilities, and confidence-guided routing. A History H row shows basketball, smartphone, and wireless mouse icons with chronological arrows. Prefix P contains instruction I and history H. Candidate branches B1, B2, and B3 show shoe, headphones, and hat icons. Icons represent item texts rather than image inputs. Each candidate branch ends with a fixed EOS input token. An eight-by-eight token attention matrix shows two schematic tokens in each of P, B1, B2, and B3, with rows attending to columns. Filled cells are visible and blank cells are blocked. The prefix and each branch have causal lower-triangular attention including the diagonal. Every branch can attend to all prefix tokens, and attention to other branches is blocked. Prefix tokens cannot attend to candidate branches. The two-token groups are illustrative and do not specify actual sequence lengths. Branch positions match independent encoding. A frozen LLM backbone with trainable LoRA adapters produces the history-conditioned EOS states r1, r2, and r3. The shared decision head applies layer normalization and a scalar projection to each state, producing one score per candidate item. Probability rows c1, c2, and c3 show bars, corresponding item icons, and illustrative values 0.72, 0.18, and 0.10. The recommendation box shows the selected-item equation and the shoe icon. Confidence at or above gamma recommends the selected item, while lower confidence routes to the larger 4B model. Dashed training paths update the shared head and LoRA at temperature one. Post-hoc calibration fits T* on D_T with model parameters frozen. The routing threshold is fixed on disjoint D_tau. The bottom legend identifies held-out data, token blocks, frozen or fixed components, trainable parameters, data flow, and loss or calibration paths.}
  \label{fig:framework}
\end{figure*}

We present Direct Item Choice Estimation (DICE), which learns a categorical distribution over candidate items from LLM representations.
We focus on independent candidate scoring (DICE-I), the main method used in our efficiency, cold-start, and confidence experiments.
We first describe candidate encoding and choice learning, then explain single-forward inference with candidate-order consistency, and finally turn the selected-item probability into calibrated confidence for selective recommendation and routing.
\paragraph{Problem Formulation.}
For sequential recommendation, let $H=(h_1,\ldots,h_L)$ denote a user's chronological interaction history, with $h_t$ the item at step $t$, and $C=(c_1,\ldots,c_K)$ a presentation of the $K$ supplied candidates.
Let $\operatorname{text}(a)$ denote item $a$'s textual description.
For a recommendation example $x=(H,C)$, $y\in\{1,\ldots,K\}$ indexes the observed next item $c_y$.
Under decision-native recommendation, we model a categorical distribution over candidate items,
\begin{equation}
  p_i(x)=\Pr(y=i\mid H,C),\quad i=1,\ldots,K,
  \label{eq:decision-distribution}
\end{equation}
and return the highest-scoring candidate.
\paragraph{Overview}
Figure~\ref{fig:framework} summarizes DICE-I's candidate encoding, scoring, shared-history inference, and confidence calibration.
\subsection{Categorical Choice Learning with LLM Representations}
\label{sec:choice}
DICE learns to select the observed next item from the supplied candidates using contextual representations from a pretrained LLM.
Each candidate is encoded and scored independently, while all candidate logits are normalized jointly and optimized through one choice loss per example.
This separates candidate-specific computation from learning the choice among candidates.
We describe the independent scorer, DICE-I, below; the candidate-interaction extension, DICE-S, is presented in Appendix~\ref{sec:appendix-dice-s}.
\paragraph{History--candidate encoding}
Let $I$ denote the textual instruction. For each candidate $c_i$, DICE constructs the token sequence
\begin{equation}
  X_i=
  [\,I;\operatorname{text}(h_1);\ldots;
  \operatorname{text}(h_L);\operatorname{text}(c_i);\mathrm{EOS}\,],
  \label{eq:candidate-input}
\end{equation}
where $[\,\cdot\,;\,\cdot\,]$ denotes token concatenation.
We separate this sequence into a shared instruction--history prefix and a candidate-specific branch:
\begin{equation}
  P=[\,I;\operatorname{text}(h_1);\ldots;\operatorname{text}(h_L)\,],
  \qquad
  B_i=[\,\operatorname{text}(c_i);\mathrm{EOS}\,],
  \label{eq:prefix-branch}
\end{equation}
so that $X_i=[P;B_i]$.
History items retain their chronological order, and each branch contains only its own candidate.
A shared causal LLM with hidden dimension $d$ encodes $X_i$.
The hidden state at the final EOS token provides the history--candidate representation:
\begin{equation}
  r_i=\text{LLM}_\theta(X_i)_{\mathrm{EOS}}\in\mathbb R^d.
  \label{eq:candidate-representation}
\end{equation}
The fixed EOS input token follows both the history and candidate, so its representation is conditioned on both.
The same encoder and input construction are used for every candidate, without encoding its displayed position in the candidate list.
The sequences $X_i$ define the candidate-specific computations; Section~\ref{sec:inference} explains how they are evaluated jointly in a single forward pass.
\paragraph{Independent candidate scoring}
A decision head shared by all candidates maps each representation to a scalar logit:
\begin{equation}
  z_i^{\mathrm I}
  =w^\top\operatorname{LN}(r_i)+b,
  \label{eq:dice-i-logit}
\end{equation}
where $w\in\mathbb R^d$, $b\in\mathbb R$, and $\operatorname{LN}$ denotes layer normalization.
Each logit depends on the history $H$ and candidate $c_i$, rather than on the candidate's displayed position or the other candidates.
Sharing the encoder and head also allows the same scoring function to handle variable candidate sets.
\paragraph{Choice distribution and training}
Although logits are computed independently, the choice probabilities depend on the complete candidate set through joint normalization.
For temperature $T>0$, DICE-I defines
\begin{equation}
  p_i^{\mathrm I}(x;T)
  =
  \frac{\exp(z_i^{\mathrm I}/T)}
       {\sum_{j=1}^{K}\exp(z_j^{\mathrm I}/T)}.
  \label{eq:choice-probability}
\end{equation}
Training uses $T=1$ and minimizes one cross-entropy loss for each example:
\begin{equation}
  \mathcal L_{\mathrm{CE}}^{\mathrm I}(x,y)
  =
  -\log p_y^{\mathrm I}(x;1).
  \label{eq:choice-loss}
\end{equation}
This objective compares the observed next item with the other candidates in the same example, jointly optimizing their scores through the shared parameters.
The pretrained backbone remains frozen. Training updates only the LoRA adapters and the shared decision head.
The objective operates on candidate-choice probabilities and does not use the LLM vocabulary head.
At inference, DICE-I selects the highest-scoring candidate:
\begin{equation}
  \hat y^{\mathrm I}(x)=\arg\max_i z_i^{\mathrm I},\qquad
  \hat c^{\mathrm I}(x)=c_{\hat y^{\mathrm I}(x)}.
  \label{eq:choice-prediction}
\end{equation}
The output therefore identifies a supplied item directly, without answer decoding or text matching.
Since neither the encoder nor the head uses item-specific output parameters, an item unseen during recommendation training can be scored from its text using the same procedure.
Section~\ref{sec:calibration} calibrates the resulting choice probabilities for use as decision confidence.
\subsection{Single-Forward Inference with Order Consistency}
\label{sec:inference}
The candidate sequences $X_i=[P;B_i]$ share the same instruction--history prefix $P$.
Encoding them independently recomputes this prefix for every candidate, even when the sequences are processed in one batch.
As illustrated in Figure~\ref{fig:framework}, DICE-I avoids this redundancy by sharing the prefix across isolated candidate branches and computing all candidate representations in a single LLM forward pass.
\paragraph{Shared-prefix inference}
DICE-I packs each example into one input,
\begin{equation}
  X_{\mathrm{tree}}=[P;B_1;\ldots;B_K],
  \label{eq:packed-input}
\end{equation}
and controls information flow with a tree-structured attention mask, following the computation pattern used in tree-based inference~\cite{miao2024specinfer,cai2024medusa}.
Prefix tokens retain standard causal attention.
Each branch token attends to the shared prefix, itself, and preceding tokens in its own branch, while attention to all other candidate branches is blocked.
Attention isolation alone is insufficient to reproduce independent encoding: the position indices must also match.
Each branch therefore uses the same position indices as in its independent sequence $X_i=[P;B_i]$, continuing from the end of the shared prefix.
The branch's physical location in the packed input does not determine its position indices.
Consequently, each candidate is evaluated with the same visible tokens and position indices as its independent counterpart.
Writing $r_i^{\mathrm{tree}}$ and $r_i^{\mathrm{ind}}$ for its tree and independently encoded EOS representations, respectively, in theory we have
\begin{equation}
  r_i^{\mathrm{tree}}=r_i^{\mathrm{ind}},\qquad i=1,\ldots,K.
  \label{eq:tree-equivalence}
\end{equation}
The shared prefix is computed once, and all $K$ EOS representations are obtained in one forward pass.
As shown in Figure~\ref{fig:framework}, the shared decision head then maps these representations to candidate logits, which are normalized into the choice distribution in Eq.~\eqref{eq:choice-probability}.
DICE-I selects the highest-scoring candidate without autoregressive answer decoding.
\paragraph{Candidate-order consistency}
Recall that $H$ denotes the chronological history and $C=(c_1,\ldots,c_K)$ denotes the candidate presentation.
Let $\pi$ be a permutation of $\{1,\ldots,K\}$, and define
\begin{equation}
  C^\pi=(c_{\pi(1)},\ldots,c_{\pi(K)}).
\end{equation}
The history $H$ remains unchanged.
Write $z_i^{\mathrm I}(H,C)$ for the DICE-I logit of the candidate at position $i$ in $C$, and $p_i^{\mathrm I}(H,C;T)$ for its choice probability at temperature $T>0$.
The latter is the notation $p_i^{\mathrm I}(x;T)$ from Section~\ref{sec:choice} with the input $x=(H,C)$ written explicitly.
Candidate isolation and shared positional framing ensure that each item is encoded with the same history and branch positions regardless of its displayed position.
The shared decision head then applies the same scoring function to every candidate.
Consequently, in exact arithmetic, the logits satisfy
\begin{equation}
  z_i^{\mathrm I}(H,C^\pi)=z_{\pi(i)}^{\mathrm I}(H,C),
  \qquad i=1,\ldots,K.
  \label{eq:permutation}
\end{equation}
The softmax denominator is unchanged by permutation, so
\begin{equation}
  p_i^{\mathrm I}(H,C^\pi;T)=p_{\pi(i)}^{\mathrm I}(H,C;T).
  \label{eq:probability-permutation}
\end{equation}
These relations express permutation equivariance: reordering candidates permutes the score and probability vectors without changing the values associated with each item.
If the maximum logit is unique, the selected item identity is unchanged.
Exact ties are resolved by a fixed item-identifier rule independent of presentation order.
Shared-prefix inference therefore reduces repeated computation while preserving the candidate-specific scoring defined in Section~\ref{sec:choice}.
Efficiency comes from sharing the prefix, whereas order consistency comes from candidate isolation, shared positional framing, and shared scoring parameters.
\subsection{Calibrated Decision Confidence}
\label{sec:calibration}
The choice distribution provides both a recommended item and a probability associated with that choice.
We use this probability as the basis for decision confidence, distinguishing two properties: \emph{discrimination}, which measures whether higher confidence identifies more correct recommendations, and \emph{calibration}, which measures whether confidence values agree with empirical accuracy.
Raw softmax probabilities may not reflect the empirical accuracy of the selected items~\cite{guo2017calibration}.
\paragraph{Temperature scaling}
After training, we freeze the backbone, adapters, and decision head.
A single positive temperature is fitted on a held-out development set $D_T$:
\begin{equation}
  T^*
  =
  \arg\min_{T>0}
  \sum_{(x,y)\in D_T}
  -\log p_y^{\mathrm I}(x;T),
  \label{eq:temperature-fit}
\end{equation}
where $x=(H,C)$ is an example, $y$ indexes its observed next item, and $p_y^{\mathrm I}(x;T)$ is the choice probability defined in Section~\ref{sec:choice}.
Temperature fitting adjusts the probability scale without updating the model parameters.
Because $T^*>0$, it preserves candidate-logit ordering and the selected item.
We define decision confidence as the calibrated probability of the selected candidate:
\begin{equation}
  s(x)
  =
  p_{\hat y^{\mathrm I}(x)}^{\mathrm I}(x;T^*),
  \label{eq:confidence}
\end{equation}
where $\hat y^{\mathrm I}(x)$ is the predicted candidate index.
Thus, $s(x)$ estimates the probability that the recommendation matches the observed next item.
Section~\ref{sec:rq3} evaluates its discrimination and calibration separately.
For selective recommendation, we retain examples whose recommendation confidence is at or above a threshold selected on a separate development set $D_\tau$, disjoint from $D_T$.
Thresholds are frozen before testing; the selection criteria and evaluation protocol are described in Appendix~\ref{sec:appendix-selective}.
\paragraph{Confidence-guided routing}
Confidence also guides computation. A smaller DICE model handles each example, deferring to a larger model when confidence is low.
Both models receive the same history and candidate set.
Let $\hat y_{\mathrm{small}}(x)$ and $\hat y_{\mathrm{large}}(x)$ denote their predicted candidate indices, and let $s_{\mathrm{small}}(x)$ denote the smaller model's calibrated decision confidence.
For a routing threshold $\gamma$, the cascade returns
\begin{equation}
  \hat y_{\mathrm{cas}}(x)
  =
  \begin{cases}
    \hat y_{\mathrm{small}}(x),
      & s_{\mathrm{small}}(x)\ge\gamma,\\
    \hat y_{\mathrm{large}}(x),
      & s_{\mathrm{small}}(x)<\gamma.
  \end{cases}
  \label{eq:routing}
\end{equation}
Here, $\hat y_{\mathrm{cas}}(x)$ is the candidate index selected by the cascade.
The threshold $\gamma$ is selected on $D_\tau$ and fixed before testing.
Figure~\ref{fig:framework} illustrates this cascade, with $s(x)=s_{\mathrm{small}}(x)$ and a 4B model handling examples below the confidence threshold.
Routing requires no separately trained router: the item-choice distribution supplies both the recommendation and the confidence used to allocate additional computation.

\section{Experiments}
\label{sec:experiments}
We evaluate DICE through three research questions:
\begin{itemize}[leftmargin=*]
  \item \textbf{RQ1: How effective is DICE under standard and item cold-start recommendation settings?}
  \item \textbf{RQ2: Does DICE ensure output validity and candidate-order
  consistency with efficient inference?}
  \item \textbf{RQ3: Is DICE's decision confidence informative, calibrated, and actionable?}
\end{itemize}
\begin{table*}[t]
  \caption{Mean recommendation performance under candidate scoring. N@3 denotes NDCG@3. Best results are \textbf{bold}; shaded rows denote DICE-I and DICE-S.}
  \label{tab:main}
  \centering
  \renewcommand{\arraystretch}{0.92}
  \setlength{\tabcolsep}{2.5pt}
  \definecolor{maintableblue}{RGB}{239,246,251}
  \shrinktofit{0.95\linewidth}{%
  \begin{tabular}{l|cccc|cccc|cccc}
    \toprule
    & \multicolumn{4}{c|}{\textbf{Goodreads}}
      & \multicolumn{4}{c|}{\textbf{LastFM}}
      & \multicolumn{4}{c}{\textbf{Amazon Industrial}} \\
    \textbf{Method}
      & \textbf{HR@1} & \textbf{HR@3} & \textbf{HR@5} & \textbf{N@3}
      & \textbf{HR@1} & \textbf{HR@3} & \textbf{HR@5} & \textbf{N@3}
      & \textbf{HR@1} & \textbf{HR@3} & \textbf{HR@5} & \textbf{N@3} \\
    \midrule
    GRU4Rec~\cite{hidasi2016gru4rec}
      & 0.3870 & 0.5482 & 0.6462 & 0.4785 & 0.2617 & 0.4389 & 0.5411 & 0.3617 & 0.2500 & 0.6046 & 0.7824 & 0.4538 \\
    Caser~\cite{tang2018caser}
      & 0.4169 & 0.5598 & 0.6578 & 0.4977 & 0.2233 & 0.4076 & 0.5166 & 0.3302 & 0.2287 & 0.5808 & 0.7652 & 0.4308 \\
    SASRec~\cite{kang2018sasrec}
      & 0.3588 & 0.5365 & 0.6379 & 0.4594 & 0.2233 & 0.3988 & 0.5074 & 0.3264 & 0.2237 & 0.5688 & 0.7567 & 0.4216 \\
    \midrule
    Base (Qwen3)~\cite{qwen2025qwen3}
      & 0.0532 & 0.1777 & 0.3090 & 0.1244 & 0.0465 & 0.1359 & 0.2373 & 0.0971 & 0.0536 & 0.1524 & 0.2496 & 0.1096 \\
    SFT~\cite{ding2025napo}
      & 0.3522 & 0.5166 & 0.6163 & 0.4457 & 0.2717 & 0.4762 & 0.5820 & 0.3901 & 0.1881 & 0.4609 & 0.6019 & 0.3466 \\
    NAPO~\cite{ding2025napo}
      & 0.3605 & 0.5764 & 0.6728 & 0.4837 & 0.3467 & 0.5503 & 0.6553 & 0.4656 & 0.2343 & 0.4802 & 0.6155 & 0.3765 \\
    S-DPO~\cite{chen2024sdpo}
      & 0.3887 & 0.5698 & 0.6794 & 0.4934 & 0.3343 & 0.5375 & 0.6501 & 0.4529 & 0.3174 & 0.5456 & 0.6764 & 0.4489 \\
    \midrule
    LlamaRec~\cite{yue2023llamarec}
      & 0.4718 & 0.7093 & 0.8123 & 0.6105 & 0.5960 & 0.7936 & 0.8609 & 0.7114 & 0.3650 & 0.6201 & 0.7568 & 0.5127 \\
    InvariRank~\cite{bito2026invarirank}
      & 0.4219 & 0.6927 & 0.8073 & 0.5788 & 0.5896 & 0.8016 & 0.8705 & 0.7142 & 0.3624 & 0.6238 & 0.7603 & 0.5137 \\
    \midrule
    RE$^2$-SFT+GRPO~\cite{re2code}
      & 0.0664 & 0.1495 & 0.2625 & 0.1138 & 0.1459 & 0.2613 & 0.3627 & 0.2120 & 0.2209 & 0.3987 & 0.5033 & 0.3237 \\
    SelfDR-style student~\cite{jiang2026selfdr}
      & 0.4402 & 0.6213 & 0.6927 & 0.5464 & 0.3968 & 0.5824 & 0.6754 & 0.5048 & 0.3533 & 0.6055 & 0.7341 & 0.4993 \\
    \midrule\midrule
    \cellcolor{maintableblue}DICE-I
      & \cellcolor{maintableblue}0.5332 & \cellcolor{maintableblue}\textbf{0.7525} & \cellcolor{maintableblue}\textbf{0.8588} & \cellcolor{maintableblue}\textbf{0.6609} & \cellcolor{maintableblue}\textbf{0.6160} & \cellcolor{maintableblue}0.8172 & \cellcolor{maintableblue}0.8898 & \cellcolor{maintableblue}0.7334 & \cellcolor{maintableblue}0.3767 & \cellcolor{maintableblue}0.6432 & \cellcolor{maintableblue}0.7808 & \cellcolor{maintableblue}0.5311 \\
    \cellcolor{maintableblue}DICE-S
      & \cellcolor{maintableblue}\textbf{0.5349} & \cellcolor{maintableblue}0.7425 & \cellcolor{maintableblue}0.8472 & \cellcolor{maintableblue}0.6552 & \cellcolor{maintableblue}0.6096 & \cellcolor{maintableblue}\textbf{0.8192} & \cellcolor{maintableblue}\textbf{0.8958} & \cellcolor{maintableblue}\textbf{0.7335} & \cellcolor{maintableblue}\textbf{0.3834} & \cellcolor{maintableblue}\textbf{0.6502} & \cellcolor{maintableblue}\textbf{0.7865} & \cellcolor{maintableblue}\textbf{0.5380} \\
    \bottomrule
  \end{tabular}}
\end{table*}
\subsection{Experimental Setup}
\label{sec:setup}
\paragraph{Datasets.}
We evaluate sequential recommendation for books, music, and industrial products.
Goodreads\footnote{\url{https://www.goodreads.com/}} provides book interactions and titles;
LastFM\footnote{\url{https://grouplens.org/datasets/hetrec-2011/}} provides listening histories and artist names;
Amazon Reviews 2023 Industrial\_and\_Scientific (Amazon Industrial)\footnote{\url{https://amazon-reviews-2023.github.io/main.html}} provides product interactions and titles.
Dataset composition, preprocessing, and candidate construction are detailed
in Appendix~\ref{sec:appendix-datasets}.
The main comparison uses Qwen3-0.6B and $K=20$ candidates.
\paragraph{Baselines.}
We compare three baseline families:
\textit{Traditional sequential recommenders.}
GRU4Rec~\cite{hidasi2016gru4rec} models interaction sequences with recurrent
networks; Caser~\cite{tang2018caser} extracts sequential patterns with
convolutional filters; SASRec~\cite{kang2018sasrec} captures historical item
dependencies with self-attention. These ID-based methods model interaction
histories without an LLM.
\textit{Generation-based LLM recommenders.}
Base~\cite{qwen2025qwen3} evaluates Qwen3 without recommendation training.
SFT~\cite{ding2025napo} learns to generate next-item names.
S-DPO~\cite{chen2024sdpo} optimizes positive--negative item preferences,
while NAPO~\cite{ding2025napo} shares negatives and adjusts reward margins
in preference training. RE$^2$-SFT+GRPO~\cite{re2code} learns to reason before
recommending through supervised fine-tuning and reinforcement learning.
SelfDR~\cite{jiang2026selfdr} distills a rationale-informed teacher into a
student that directly generates a candidate letter without a rationale at inference.
\textit{Direct-ranking LLM recommenders.}
LlamaRec~\cite{yue2023llamarec} uses a verbalizer to map candidate-letter
logits to item scores. InvariRank~\cite{bito2026invarirank} isolates candidate
attention and shares positional framing to stabilize scores under reordering.
\paragraph{Evaluation.}
The main table evaluates candidate ranking, selecting each method's
highest-scoring candidate. Generated answers are evaluated separately by
mapping decoded outputs to candidates; unmatched or ambiguous answers
count as incorrect.
HR@1 measures top-one accuracy. HR@$k$ and NDCG@$k$ assess ranking
quality. ValidRatio is the fraction of outputs that identify a unique
candidate.
Decision confidence is the probability assigned to the selected item.
Expected calibration error (ECE) is the example-weighted absolute gap
between mean confidence and top-one accuracy over ten equal-width
confidence bins~\cite{guo2017calibration}. Lower ECE indicates better
confidence--accuracy agreement.\footnote{Scoring definitions
and the ECE formula appear in Appendices~\ref{sec:appendix-readouts}
and~\ref{sec:appendix-selective}, respectively.}
For all comparisons, we report arithmetic means of per-run empirical metrics
over three random seeds.
\subsection{RQ1: Recommendation Quality and Item Cold-Start Performance}
\label{sec:results}
We compare DICE with baselines across three datasets (Table~\ref{tab:main}), then examine backbone scaling (Table~\ref{tab:model-scale}) and item cold-start performance (Table~\ref{tab:cold-start}).
\textbf{DICE improves top-one accuracy and ranking quality.} Table~\ref{tab:main} shows DICE-I exceeding the baselines in
HR@1, HR@3, and NDCG@3, indicating better top-one selection and target
placement near the top of the ranking. Its HR@5 also
exceeds the baselines on Goodreads
and LastFM; on Amazon Industrial, GRU4Rec is slightly higher
(0.7824 vs. 0.7808).
Both DICE variants occupy the top two positions across the metrics
on Goodreads and LastFM. DICE-S leads all four metrics on Amazon Industrial,
indicating that candidate interactions can improve ranking, although their
benefits over DICE-I are mixed on Goodreads and LastFM.
\indent\textbf{A simple decision interface outperforms direct-ranking baselines.}
DICE-I exceeds LlamaRec and InvariRank
across all four metrics in Table~\ref{tab:main}.
Compared with LlamaRec, the strongest HR@1 baseline, DICE-I achieves
absolute HR@1 gains of +6.14\%, +2.00\%, and +1.17\% on Goodreads,
LastFM, and Amazon Industrial, respectively.
These results support effective ranking with a shared scoring head and
categorical choice learning.\footnote{Additional comparisons with frozen
text representations and an ID-only recommender appear in
Appendix~\ref{sec:appendix-representation}.}
\indent\textbf{DICE selects items accurately without test-time reasoning.}
Table~\ref{tab:main} shows that DICE-I outperforms both the SelfDR-style
student, which generates a candidate letter without a rationale,
and RE$^2$-SFT+GRPO, which generates a rationale before answering.
These results support effective next-item selection without generating
reasoning or answer text at inference.
The SelfDR-style student also outperforms SFT, S-DPO, and NAPO,
suggesting that reasoning-informed supervision can help a model that
recommends directly. The separate ranking--generation comparison
appears in Table~\ref{tab:interface-diagnostic}.
\begin{table}[!t]
  \caption{Mean DICE-I performance across Qwen3 backbone sizes.
  N@5 denotes NDCG@5; higher is better.}
  \label{tab:model-scale}
  \centering\small
  \renewcommand{\arraystretch}{1.08}
  \setlength{\tabcolsep}{2pt}
  \begin{tabular*}{\columnwidth}{@{\extracolsep{\fill}}l*{6}{r}@{}}
    \toprule
    & \multicolumn{3}{c}{\textbf{Goodreads}} & \multicolumn{3}{c}{\textbf{LastFM}} \\
    \cmidrule(lr){2-4}\cmidrule(l){5-7}
    \textbf{Size} & \textbf{HR@1} & \textbf{HR@3} & \textbf{N@5} & \textbf{HR@1} & \textbf{HR@3} & \textbf{N@5} \\
    \midrule
    0.6B & 0.5332 & 0.7525 & 0.7048 & 0.6160 & 0.8172 & 0.7633 \\
    1.7B & \textbf{0.5465} & 0.7641 & 0.7090 & 0.6509 & 0.8529 & 0.7912 \\
    4B & \textbf{0.5465} & \textbf{0.7741} & \textbf{0.7212}
       & \textbf{0.7018} & \textbf{0.8926} & \textbf{0.8368} \\
    \bottomrule
  \end{tabular*}
\end{table}
\indent\textbf{Larger backbones help, especially on LastFM.}
Table~\ref{tab:model-scale} shows LastFM HR@1 rising from 0.6160 to
0.7018 as the backbone grows, alongside improvements in HR@3 and NDCG@5.
Goodreads HR@1 reaches 0.5465 at 1.7B and then plateaus, while HR@3 and
NDCG@5 continue to improve. Overall, scaling yields larger gains on LastFM
and improves ranking beyond the first position on Goodreads.
The common training protocol appears in Appendix~\ref{sec:appendix-training}.
\begin{table}[!t]
  \caption{Mean cold-target HR@1 with 10 cold and 10 warm candidates.
  Best results are bold; protocols appear in Appendix~\ref{sec:appendix-cold-start}.}
  \label{tab:cold-start}
  \centering\small
  \renewcommand{\arraystretch}{1.04}
  \setlength{\tabcolsep}{5pt}
  \begin{tabular}{lrrr}
    \toprule
    \textbf{Method} & \textbf{Goodreads} & \textbf{LastFM} & \makecell{\textbf{Amazon}\\\textbf{Industrial}} \\
    \midrule
    SASRec & 0.0909 & 0.0855 & 0.0750 \\
    SFT & \textbf{0.2562} & 0.1842 & 0.1490 \\
    S-DPO & \textbf{0.2562} & 0.1952 & 0.2153 \\
    \midrule
    \rowcolor{oursrowblue}
    DICE-I & 0.2397 & \textbf{0.2303} & \textbf{0.2495} \\
    \bottomrule
  \end{tabular}
\end{table}
\indent\textbf{DICE supports item cold-start recommendation.}
In Table~\ref{tab:cold-start}, DICE-I achieves HR@1 of 0.2397 on
Goodreads and 0.2303 on LastFM, exceeding SASRec's 0.0909 and 0.0855.
Its relative advantage among LLM methods varies: SFT and S-DPO both
reach 0.2562 on Goodreads, while DICE-I exceeds their 0.1842 and 0.1952
on LastFM. On Amazon Industrial, DICE-I reaches 0.2495, ahead of
S-DPO (0.2153), SFT (0.1490), and SASRec (0.0750).
These results support effective selection of items unseen
during recommendation training, with the strongest method varying across datasets.
\subsection{RQ2: Structural Properties and Inference Efficiency}
\label{sec:rq2}
\begin{figure}[!htbp]
  \centering
  \begin{subfigure}[b]{0.49\columnwidth}
    \centering
    \includegraphics[width=\linewidth]{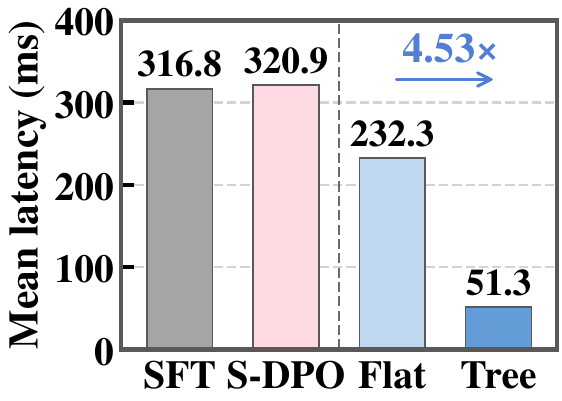}
    \caption{Goodreads.}
  \end{subfigure}\hfill
  \begin{subfigure}[b]{0.49\columnwidth}
    \centering
    \includegraphics[width=\linewidth]{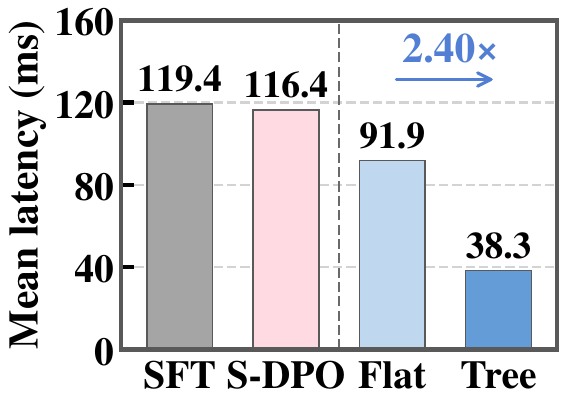}
    \caption{LastFM.}
  \end{subfigure}
  \caption{Mean end-to-end latency ($K=20$, 200 examples per run and dataset).
  All paths include tokenization. Flat/Tree denote DICE-I layouts;
  arrows give flat-to-tree speedups. Lower is better.}
  \Description{Two bar charts show mean latency for SFT generation, S-DPO generation, DICE-I flat, and DICE-I tree. Goodreads values are 316.81, 320.93, 232.34, and 51.27 milliseconds; LastFM values are 119.41, 116.44, 91.86, and 38.34 milliseconds. A dashed divider separates generation baselines from DICE-I scoring paths. Tree accelerates the matched flat path by 4.53 and 2.40 times, respectively.}
  \label{fig:latency}
\end{figure}

\begin{figure}[!htbp]
  \centering
  \includegraphics[width=\columnwidth]{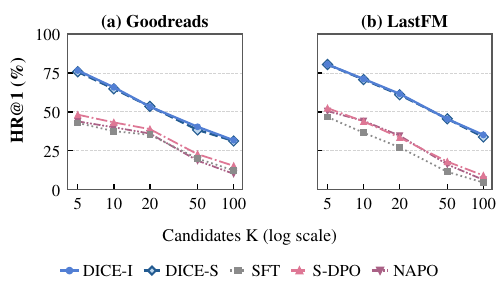}
  \caption{Mean HR@1 across candidate counts ($K$ includes the target).
  The horizontal axis is logarithmic.}
  \Description{Goodreads and LastFM panels compare DICE-I, DICE-S, SFT,
  S-DPO, and NAPO at 5, 10, 20, 50, and 100 candidates. All methods decline
  as candidates increase, while both DICE variants exceed the three
  compared baselines.}
  \label{fig:rq2-candidate-sensitivity}
\end{figure}

We compare candidate scoring with answer generation (Table~\ref{tab:interface-diagnostic}) and assess candidate-order consistency (Table~\ref{tab:permutation}), followed by analyses of inference latency and candidate-set size.
\indent\textbf{Candidate scoring improves selection over answer generation.}
In Table~\ref{tab:interface-diagnostic}, candidate scoring improves
SFT, S-DPO, and NAPO HR@1 by
2.82--8.81\% on Goodreads and 4.09--6.41\% on LastFM (absolute gains).
RE$^2$ achieves nearly identical HR@1 under ranking and generation,
indicating that the accuracy benefit
of ranking by candidate-name likelihood varies across methods. These results support
candidate scoring as a practical recommendation interface.
A separate single-checkpoint SFT diagnostic finds that 16.11\% and 7.98\%
of generated answers fail to identify a unique candidate item on Goodreads
and LastFM, respectively.
Direct candidate selection avoids unmatched or ambiguous answers.
DICE adopts this selection interface and additionally provides
order-consistent decisions and calibrated confidence.
\begin{table}[!htbp]
  \caption{Mean HR@1: ranking (Rank) versus generation (Gen.).}
  \label{tab:interface-diagnostic}
  \centering\small
  \setlength{\tabcolsep}{3pt}
  \renewcommand{\arraystretch}{1.08}
  \begin{tabular*}{\columnwidth}{@{\extracolsep{\fill}}lrrrr}
    \toprule
    & \multicolumn{2}{c}{\textbf{Goodreads}} & \multicolumn{2}{c}{\textbf{LastFM}} \\
    \cmidrule(lr){2-3}\cmidrule(lr){4-5}
    \textbf{Method} & \textbf{Rank} & \textbf{Gen.} & \textbf{Rank} & \textbf{Gen.} \\
    \midrule
    Base & 0.0532 & 0.0000 & 0.0465 & 0.0000 \\
    SFT & 0.3522 & 0.2807 & 0.2717 & 0.2076 \\
    S-DPO & 0.3887 & 0.3605 & 0.3343 & 0.2934 \\
    NAPO & 0.3605 & 0.2724 & 0.3467 & 0.2854 \\
    RE$^2$-SFT+GRPO & 0.0664 & 0.0664 & 0.1459 & 0.1455 \\
    \bottomrule
  \end{tabular*}
\end{table}
\begin{table}[!htbp]
  \caption{Mean candidate-order FlipRate (lower is better):
  64 examples $\times$ 20 reorderings per run, method, and dataset.
  SFT/S-DPO generate answers.}
  \label{tab:permutation}
  \centering\small
  \setlength{\tabcolsep}{3pt}
  \begin{tabularx}{\columnwidth}{@{}l*{4}{>{\centering\arraybackslash}X}@{}}
    \toprule
    & \multicolumn{2}{c}{Generation-based} & \multicolumn{2}{c}{DICE} \\
    \cmidrule(lr){2-3}\cmidrule(l){4-5}
    \textbf{Dataset} & SFT & S-DPO & DICE-I & DICE-S \\
    \midrule
    Goodreads & 0.3180 & 0.4500 & \textbf{0.0000} & \textbf{0.0000} \\
    LastFM & 0.2984 & 0.3125 & \textbf{0.0000} & \textbf{0.0000} \\
    \bottomrule
  \end{tabularx}
\end{table}
\indent\textbf{DICE preserves output validity and choices under candidate reordering.}
Neither DICE variant changes its selected item, logit, or probability
under the tested reorderings, whereas SFT and S-DPO exhibit choice or
validity changes (Table~\ref{tab:permutation}).
Under candidate reordering, SFT achieves ValidRatio of 80.55\%--92.58\%
across the two datasets, whereas DICE maintains 100\% by selecting
directly from the supplied candidates
(Appendix~\ref{sec:appendix-order}, Table~\ref{tab:reordering-validity}).
DICE-I uses the same independent scoring function for every candidate,
so reordering only permutes its logits and probabilities
(Eqs.~\eqref{eq:permutation}--\eqref{eq:probability-permutation}). The DICE-S extension preserves this property
through a permutation-equivariant interaction module
(Appendix~\ref{sec:appendix-dice-s}). \footnote{Appendix~\ref{sec:appendix-order}
defines the protocol and FlipRate.}
\indent\textbf{DICE-I reduces latency versus generation and flat encoding.}
Figure~\ref{fig:latency} shows tree latency of 51.27 and 38.34 ms on Goodreads
and LastFM, respectively,
versus SFT's 316.81/119.41 ms and S-DPO's 320.93/116.44 ms.
The end-to-end speedup over these answer-generation paths is 6.18--6.26$\times$ on Goodreads and 3.04--3.11$\times$ on LastFM.
Compared with DICE-I flat encoding, tree inference is 4.53$\times$ faster
on Goodreads and 2.40$\times$ faster on LastFM.
These gains show the benefit of sharing history computation.
Flat and tree select the same item on every checked example, so the
runtime reduction preserves the evaluated decisions.\footnote{Timing protocols appear in Appendices~\ref{sec:appendix-latency} and~\ref{sec:appendix-tree}.}
\indent\textbf{DICE retains an advantage across candidate counts.}
Both DICE variants exceed SFT, S-DPO, and NAPO at every tested candidate
count (Figure~\ref{fig:rq2-candidate-sensitivity}), although all methods'
HR@1 declines as the choice set grows. At $K=100$, DICE-I exceeds the
strongest compared baseline, S-DPO, with absolute HR@1 gains of 16.45\%
on Goodreads and 26.09\% on LastFM. The advantage thus persists with more competing
items. DICE-I and DICE-S show similar trends without a consistent winner
(protocol in Appendix~\ref{sec:appendix-candidate-sensitivity}).
\subsection{RQ3: Decision Confidence and Applications}
\label{sec:rq3}
\begin{figure}[!t]
  \centering
  \begin{subfigure}[b]{0.49\columnwidth}
    \centering
    \includegraphics[width=\linewidth]{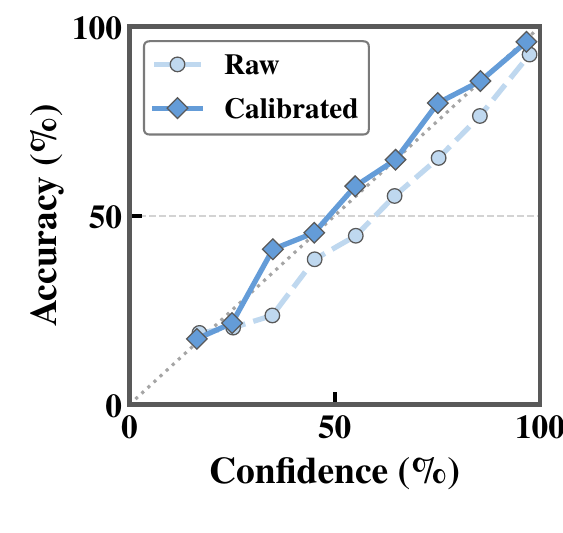}
    \caption{Calibration, LastFM.}
    \label{fig:rq3-calibration}
  \end{subfigure}\hfill
  \begin{subfigure}[b]{0.49\columnwidth}
    \centering
    \includegraphics[width=\linewidth]{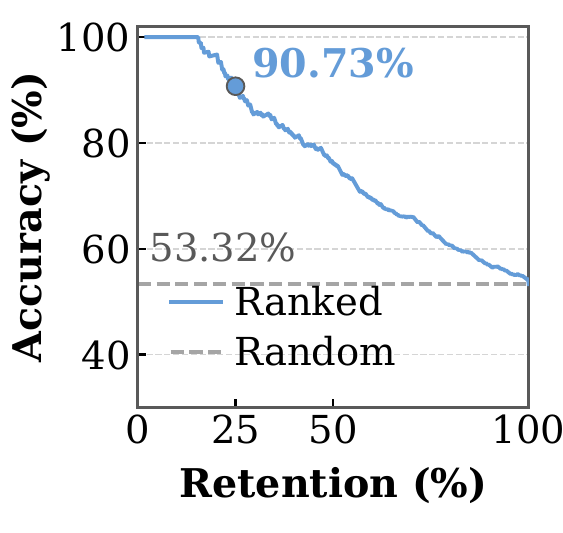}
    \caption{Selective, Goodreads.}
    \label{fig:rq3-dice-coverage}
  \end{subfigure}
  \caption{DICE-I confidence. (a) LastFM calibration before and after
  per-run temperature scaling. (b) Mean Goodreads accuracy versus
  retention, ranking test confidence within each run. The dashed line
  shows random selection; the marker denotes the most confident 25\%.}
  \Description{The left panel compares raw and calibrated selected-item
  probabilities with empirical bin accuracy on LastFM. The right panel plots
  Goodreads accuracy against test retention. The most confident
  quarter achieves 90.73 percent accuracy; full retention and the
  random-retention baseline both equal 53.32 percent.}
  \label{fig:rq3-confidence}
\end{figure}

We compare confidence calibration across methods (Table~\ref{tab:rq3-dice-probability}), then examine its use in selective recommendation and confidence-guided routing.
\indent\textbf{Temperature scaling improves calibration across methods.}
Table~\ref{tab:rq3-dice-probability} shows that temperature scaling reduces
ECE for all three methods on both datasets.
Temperatures are fitted separately for each method on held-out development
data, following the protocol in Appendix~\ref{sec:appendix-selective}.
After calibration, DICE-I achieves the lowest ECE, reaching 0.0538 on
Goodreads and 0.0203 on LastFM, below both LlamaRec and InvariRank.
These results indicate closer agreement between DICE-I's decision
confidence and recommendation accuracy.
Figure~\ref{fig:rq3-calibration} illustrates its calibration before and
after temperature scaling.
\begin{table}[!t]
  \caption{Mean per-run ECE. Raw: $T=1$; calibrated:
  $T^*$ fitted per method and run on development data.
  Best per dataset and column are bold; lower is better.}
  \label{tab:rq3-dice-probability}
  \centering\small
  \begin{tabular*}{\columnwidth}{@{\extracolsep{\fill}}llrr@{}}
    \toprule
    \textbf{Dataset} & \textbf{Method} & \textbf{Raw ECE} & \textbf{Cal. ECE} \\
    \midrule
    \multirow{3}{*}{Goodreads}
      & LlamaRec & 0.1024 & 0.0631 \\
      & InvariRank & 0.0887 & 0.0584 \\
      & \textbf{DICE-I} & \textbf{0.0606} & \textbf{0.0538} \\
    \midrule
    \multirow{3}{*}{LastFM}
      & LlamaRec & 0.0918 & 0.0336 \\
      & InvariRank & 0.0836 & 0.0287 \\
      & \textbf{DICE-I} & \textbf{0.0730} & \textbf{0.0203} \\
    \bottomrule
  \end{tabular*}
\end{table}
\indent\textbf{Confidence helps identify recommendations more likely to be correct.}
Using confidence thresholds selected on development data and fixed before
testing, DICE-I retains roughly a quarter of test examples, achieving
recommendation accuracies of 88.82\% on Goodreads and 94.42\% on LastFM.
The absolute accuracy gains over the full-test values are 35.50\%
and 32.82\%, showing that the examples retained by confidence have
higher recommendation accuracy.
Figure~\ref{fig:rq3-dice-coverage} instead ranks Goodreads test examples
by confidence, showing 90.73\% accuracy at 25\% retention and 53.32\%
at full retention.
The frozen-threshold rules appear in
Appendix~\ref{sec:appendix-selective}.
\begin{figure}[!t]
  \centering
  \begin{subfigure}[b]{0.49\columnwidth}
    \centering
    \includegraphics[width=\linewidth]{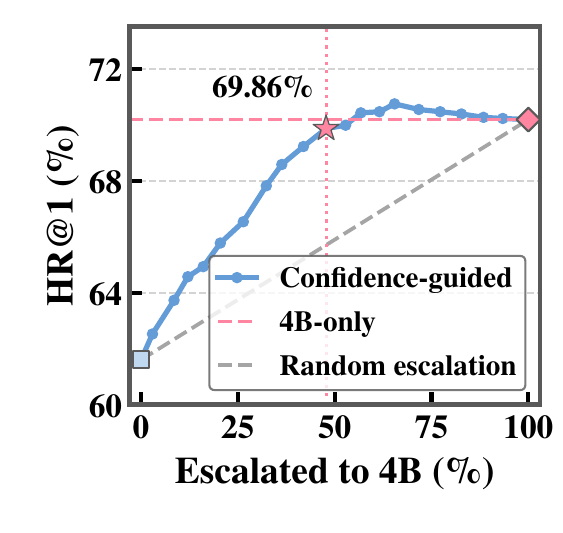}
    \caption{Recommendation quality.}
  \end{subfigure}\hfill
  \begin{subfigure}[b]{0.49\columnwidth}
    \centering
    \includegraphics[width=\linewidth]{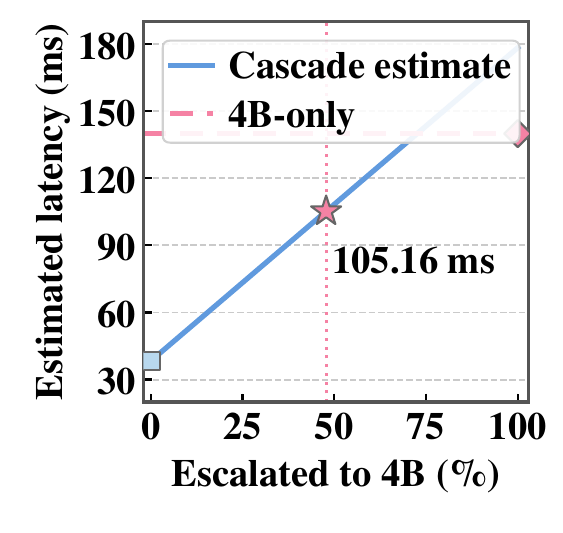}
    \caption{Computation cost.}
  \end{subfigure}
  \caption{DICE-I routing on LastFM: mean accuracy versus mean test escalation,
  with sequential latency estimated from component costs.
  Both models and the 4B-only reference use tree inference; routing overhead is excluded.}
  \Description{Two side-by-side panels show accuracy and estimated sequential
  latency against the percentage of examples escalated to the 4B DICE-I model.
  Confidence-guided routing exceeds random escalation at matched rates. At the
  highlighted 47.78 percent rate, HR@1 is 69.86 percent and
  estimated sequential latency is 105.16 milliseconds. Square and diamond markers denote
  0.6B-only and 4B-only DICE-I models. The direct 4B baseline is distinct from
  the all-escalation cascade endpoint, which runs both models.}
  \label{fig:router}
\end{figure}
\indent\textbf{Confidence-guided routing reduces estimated sequential latency.}
\label{sec:rq3-routing}
At the highlighted LastFM point in Figure~\ref{fig:router}, escalating
47.78\% of examples achieves HR@1 of 0.6986, retaining over 99.5\% of
the 4B-only model's 0.7018. Estimated sequential latency is 105.16 ms,
24.8\% below the 4B-only tree model. At the same escalation rate, the
random-routing reference is 0.6570 HR@1; confidence-guided routing yields an absolute
HR@1 gain of 4.16\%. Thus,
confidence directs additional computation toward examples that benefit
from the larger model, preserving most of its accuracy while reducing
estimated cost.\footnote{The routing and latency-estimation protocols appear in
Appendix~\ref{sec:appendix-routing}.}

\section{Conclusion and Discussion}
\label{sec:conclusion}
We introduced decision-native recommendation as a paradigm that treats valid candidate selection, candidate-order consistency, and confidence that reflects observed recommendation accuracy as joint requirements. DICE realizes these requirements through categorical choice learning with LLM representations, shared-history inference, and held-out calibration. Across three benchmarks, DICE achieves higher top-one accuracy than all evaluated baselines. Additional evaluations support selection of items unseen during recommendation training and order-consistent inference. Matched DICE-I flat--tree measurements demonstrate the computational benefit of history sharing with the scorer held fixed. In calibration evaluations on two benchmarks, DICE-I has lower reported mean ECE than both direct-ranking baselines after method-specific held-out temperature scaling. Selective recommendation and offline routing illustrate how confidence supports prediction selection and computation allocation.
Future work will extend decision-native recommendation to settings that couple candidate retrieval with item selection, including abstention when the target is absent from the retrieved pool. We will investigate adaptive calibration as candidate composition and user behavior evolve. Another direction is to extend confidence-guided routing to online serving under dynamic workloads and application-specific latency budgets. This extension will explore routing policies that account for the expected accuracy gain from escalation together with its computational and serving costs.

\clearpage%
\bibliographystyle{ACM-Reference-Format}
\bibliography{references}
\appendix
\section{Evaluation Details}
\label{sec:appendix}
\label{sec:appendix-protocol}
Anonymous repository: \url{https://anonymous.4open.science/r/dice-73B3/}.
\subsection{Datasets and Preprocessing}
\label{sec:appendix-datasets}
We follow the chronological 8:1:1 data-splitting protocol of
S-DPO and NAPO~\cite{chen2024sdpo,ding2025napo}. On Goodreads, users and items
with fewer than 20 interactions are removed~\cite{chen2024sdpo,ding2025napo}.
Each example pairs a chronological interaction history with its next observed
item. We retain histories with at least three non-padding items and do not
deduplicate histories across partitions. Table~\ref{tab:data} summarizes
dataset composition.
For the main comparison ($K=20$), each example contains its observed target
and 19 uniformly sampled negatives drawn without replacement from the item
catalog, excluding the target and history items. Candidate order is
randomized per example independently of the gold item to reduce answer-position
bias. Items are represented by book or product titles, or artist names. Evaluation examples are aligned across
methods by example ID, chronological history, candidate IDs and order,
and target item.
\begin{table}[t]
  \caption{Dataset composition.}
  \label{tab:data}
  \centering\small
  \setlength{\tabcolsep}{3pt}
  \begin{tabular*}{\columnwidth}{@{\extracolsep{\fill}}lrrr}
    \toprule
    \textbf{Dataset} & \textbf{Users} & \textbf{Items} & \textbf{Interactions} \\
    \midrule
    Goodreads & 6,031 & 4,500 & 220,100 \\
    LastFM & 1,220 & 4,606 & 73,510 \\
    Amazon 2023 Industrial & 51,000 & 25,800 & 412,900 \\
    \bottomrule
  \end{tabular*}
\end{table}
\subsection{Item Cold-Start Protocol}
\label{sec:appendix-cold-start}
The Goodreads/LastFM cold-start experiment follows the main training and candidate-ranking
protocols, with held-out-item generalization motivated by UniSRec and
Recformer~\cite{hou2022unisrec,li2023recformer}.
We hold out approximately 20\% of items as cold items and exclude them
from recommendation training. Training histories,
targets, and candidates contain only the remaining items; the chronological
partitions, method-specific training objectives, and scoring rules follow
the main experiment.
Evaluation uses shared candidate sets with unique normalized names:
each example contains 10 cold and 10 warm candidates, including the target.
Table~\ref{tab:cold-start} reports HR@1 on test examples whose targets
belong to these held-out items in Goodreads, LastFM, and Amazon Industrial.
Cold items are unseen during recommendation training, but may have appeared
in LLM pretraining.
Amazon Industrial uses normalized-title-grouped holdout and warm-only Qwen3-0.6B LoRA training
for three epochs; S-DPO starts from the warm-only SFT model.
Its SASRec boundary comparison uses shared item embeddings with
dot-product/BCE training,
selecting the epoch by warm-validation negative log-likelihood.
\subsection{Training and Baseline Implementation}
\label{sec:appendix-training}
The experimental environment provides eight NVIDIA H20 GPUs.
All training, evaluation, and timing experiments use FP32.
Methods share data partitions, histories, candidate sets and order, and
evaluation protocols. All LLM methods use Qwen3-0.6B in the main comparison;
method-specific scoring interfaces appear in Appendix~\ref{sec:appendix-readouts}.
Traditional methods use Adam with learning rate $10^{-3}$ and batch size
256. L2 coefficients are tuned over
$\{10^{-3},10^{-4},10^{-5},10^{-6},10^{-7}\}$.
DICE-I and DICE-S train LoRA~\cite{hu2021lora} adapters on attention projections
(rank 8, $\alpha=32$, dropout 0.05) and their heads, with the backbone frozen.
Optimization uses AdamW with learning rate $10^{-5}$, effective batch size
128, 5\% warm-up, and cosine learning-rate decay. DICE-I/S train for three epochs,
with epoch-3 checkpoints used for evaluation. DICE-I is the main method;
DICE-S is the candidate-interaction extension (Appendix~\ref{sec:appendix-dice-s}).
Other LLM baselines follow their original training procedures and
method-specific hyperparameters within the shared backbone, data, and
candidate setting.
DICE-I backbone sizes share the scoring head, choice loss, adaptation
strategy, and evaluation examples; computational budgets vary with model size. RQ2 reuses the RQ1 DICE-I checkpoints;
RQ3 follows the held-out calibration protocol below.
\subsection{Stability Across Random Seeds}
\label{sec:appendix-seed-stability}
Table~\ref{tab:seed-stability} reports HR@1 means and sample standard
deviations for four direct-ranking methods across three datasets.
DICE-I achieves higher mean HR@1 than LlamaRec and InvariRank on all three
datasets, with sample standard deviations no greater than 0.0017.
These standard deviations are smaller than the corresponding mean HR@1
gains over either baseline.
\begin{table}[!htbp]
  \caption{HR@1 as mean $\pm$ sample standard deviation ($n=3$).}
  \label{tab:seed-stability}
  \centering\small
  \setlength{\tabcolsep}{2pt}
  \begin{tabular*}{\columnwidth}{@{\extracolsep{\fill}}lrrr@{}}
    \toprule
    Method & Goodreads & LastFM & \makecell{Amazon\\Industrial} \\
    \midrule
    LlamaRec & $0.4718 \pm 0.0029$ & $0.5960 \pm 0.0008$ & $0.3650 \pm 0.0012$ \\
    InvariRank & $0.4219 \pm 0.0023$ & $0.5896 \pm 0.0007$ & $0.3624 \pm 0.0010$ \\
    DICE-I & $0.5332 \pm 0.0017$ & $0.6160 \pm 0.0004$ & $0.3767 \pm 0.0008$ \\
    DICE-S & $0.5349 \pm 0.0017$ & $0.6096 \pm 0.0006$ & $0.3834 \pm 0.0009$ \\
    \bottomrule
  \end{tabular*}
\end{table}
\subsection{Candidate Scoring and Generated-Answer Evaluation}
\label{sec:appendix-readouts}
\paragraph{Direct-ranking baselines.}
We follow each method's original scoring interface.
LlamaRec maps candidate-letter logits to item scores with a
verbalizer~\cite{yue2023llamarec}. Candidates use A--T.
InvariRank isolates candidate attention, shares positional framing, and
scores each candidate span by its mean token log
probability~\cite{bito2026invarirank}.
\paragraph{Generation-based baselines.}
Base, SFT, S-DPO, NAPO, and RE$^2$-SFT+GRPO rank candidates by the likelihood
of each item name, summing its token log-probabilities:
\begin{equation}
  z_i=\sum_{t=1}^{L_i}\log P_\theta(w_{i,t}\mid x,w_{i,<t}).
  \label{eq:generative-candidate-scores}
\end{equation}
Here, $w_{i,t}$ is token $t$ of the item name, $L_i$ is its token length, and the
conditioning context follows each method's prompt; RE$^2$ retains its
generated reasoning in this context.
The SelfDR student ranks candidates using candidate-letter logits at the
answer position~\cite{jiang2026selfdr}, with candidates labeled A--T.
\paragraph{Item-score interfaces.}
GRU4Rec, Caser, and SASRec use their method-specific item scores;
DICE ranks its decision-head logits directly.
Generated-answer results are reported separately from candidate scoring.
Base, SFT, S-DPO, and NAPO use greedy answer generation.
ValidRatio~\cite{ding2025napo} is computed here as the fraction of answers
uniquely matching a full candidate name after normalization; unmatched or ambiguous
answers count as incorrect. For letter-answer methods, decoded letters
are mapped to item IDs; invalid or ambiguous answers count as incorrect.
Table~\ref{tab:interface-diagnostic} evaluates both readouts on each run's
checkpoint using identical examples and candidates.
\subsection{Candidate-Order Consistency Protocol}
\label{sec:appendix-order}
Each run uses 64 fixed test examples per dataset (Goodreads and LastFM),
with 20 distinct candidate orders each: $1{,}280$ reordering inferences per
run, method, and dataset. Histories and candidate sets remain
fixed~\cite{bito2025positionbias,bito2026preferenceconsistency}.
For examples $D$ and orders $\Pi$, $\hat c(x)$ and $\hat c_\pi(x)$ denote
original and reordered selections, compared by item identity:
\begin{equation}
  \operatorname{FlipRate}=\frac{1}{|D||\Pi|}
  \sum_{x\in D}\sum_{\pi\in\Pi}
  \mathbf{1}\{\hat c_\pi(x)\ne\hat c(x)\}.
  \label{eq:fliprate}
\end{equation}
Nonunique generated answers are denoted $\bot$: transitions between
$\bot$ and a valid item count as flips, but two invalid outputs do not.
DICE logits and probabilities are aligned by
item identity before comparison with the original order; their maximum
absolute differences are zero for both variants across all 1,280 reorderings
per run and dataset.
\begin{table}[!htbp]
  \caption{Output validity under candidate reordering.
  ValidRatio is the fraction of outputs that uniquely identify a candidate.
  DICE selects directly from the supplied candidates.}
  \label{tab:reordering-validity}
  \centering\small
  \begin{tabular*}{\columnwidth}{@{\extracolsep{\fill}}lrr@{}}
    \toprule
    \textbf{Dataset} & \textbf{SFT ValidRatio} $\uparrow$ & \textbf{DICE ValidRatio} $\uparrow$ \\
    \midrule
    Goodreads & 80.55\% & \textbf{100\%} \\
    LastFM & 92.58\% & \textbf{100\%} \\
    \bottomrule
  \end{tabular*}
\end{table}
\subsection{Latency Protocol}
\label{sec:appendix-latency}
Figure~\ref{fig:latency} times four paths at $K=20$ on the same H20 GPU,
with 200 examples per run and dataset, sharing examples, histories,
candidates, and prompt construction. End-to-end timing covers tokenization,
input preparation, GPU transfer, inference, and selection, plus answer
decoding/parsing for SFT/S-DPO and packing/masks for DICE-I.
Model loading and warm-up are excluded.
Flat independently encodes the $K$ history--candidate sequences in one
batch; tree shares their prefix across isolated branches.
Both DICE-I paths use one LLM forward pass and zero decoded tokens.
Latency is averaged over examples within each run; the flat/tree protocol's
three timing rounds are separate from training runs.
The generation/tree comparison measures complete inference paths under
the same timing protocol. The matched flat/tree comparison evaluates
prefix sharing with the DICE-I model held fixed.
S-DPO was selected by development answer-generation accuracy rather than test latency.
Reasoning-baseline timings are omitted because their answer protocols and serving stacks differ.
\subsection{Flat--Tree Inference Comparison}
\label{sec:appendix-tree}
Using RQ1 DICE-I checkpoints, flat and tree achieve mean item-choice
agreement of 100\% on 100 fixed examples per run and
dataset (Goodreads/LastFM). Small probability differences remain;
this check differs from candidate reordering (Appendix~\ref{sec:appendix-order}).
\subsection{Candidate-Count Sensitivity}
\label{sec:appendix-candidate-sensitivity}
Figure~\ref{fig:rq2-candidate-sensitivity} uses
$K\in\{5,10,20,50,100\}$ and the same complete test sets, with nested candidate sets that include
the observed target and preserve shared candidates' relative order.
DICE-I/S use Qwen3-0.6B with LoRA. Generation-based baselines rank candidate-name
completion log likelihoods rather than generated answers.
HR@1 is calculated separately at each candidate count.
\subsection{Decision-Confidence Evaluation Protocol}
\label{sec:appendix-selective}
Calibration and selection use frozen 0.6B DICE-I checkpoints at
$K=20$ on the complete Goodreads and LastFM test sets.
Each run fits temperature on $D_T$ and selects thresholds on disjoint $D_\tau$.
Table~\ref{tab:rq3-dice-probability} also applies temperature scaling to
LlamaRec and InvariRank, fitting $T^*$ per method and run on $D_T$.
Test ECE is computed per run before averaging;
probabilities are not averaged before metric calculation.
Here $p_i(x)=p_i(x;T)$, with raw $T=1$
or calibrated $T=T^*$.
Expected calibration error (ECE) bins selected-item probabilities
$\max_i p_i(x)$ into ten equal-width groups $\mathcal B_m$
~\cite{guo2017calibration}. For $N=|D|$ test examples,
\begin{equation}
  \operatorname{ECE}=\sum_{m:\,|\mathcal B_m|>0}\frac{|\mathcal B_m|}{N}
  \left|\operatorname{acc}(\mathcal B_m)-\operatorname{conf}(\mathcal B_m)\right|.
  \label{eq:ece}
\end{equation}
Here, $\operatorname{acc}(\mathcal B_m)$ is the fraction of examples in
the bin whose selected item is correct, and
$\operatorname{conf}(\mathcal B_m)=|\mathcal B_m|^{-1}\sum_{x\in \mathcal B_m}\max_i p_i(x)$ is its mean
selected-item probability. For calibrated predictions, $\max_i p_i(x)$ is
the decision confidence $s(x)$ (Equation~\ref{eq:confidence}).
Figure~\ref{fig:rq3-calibration} bins raw and calibrated selected-item
probabilities separately, omitting bins
with fewer than ten test examples; ECE includes all examples.
\paragraph{Selective recommendation.}
Confidence thresholding retains examples with $s(x)\ge\tau$, leaving their
candidate sets unchanged. Accuracy is the fraction of retained examples
with a correct item choice; coverage is the retained fraction of all examples.
Empty-set accuracy is undefined. Thresholds are chosen on development data,
without test labels.
For target retention $\alpha$, $\hat\tau_\alpha$ is the confidence at rank
$\lceil\alpha|D_\tau|\rceil$ on $D_\tau$, sorted by decreasing confidence.
Retaining all ties, the frozen threshold defines
\begin{equation}
  \mathcal S_\alpha(D)=\{x\in D:s(x)\ge\hat\tau_\alpha\}.
  \label{eq:fixed-service}
\end{equation}
Thresholds are frozen per run before averaging test accuracy and coverage;
test coverage need not equal $\alpha$. Figure~\ref{fig:rq3-dice-coverage}
instead averages accuracy from each run's test confidence ranking at common
retention targets.
The main-text selective results use different development criteria.
Goodreads selects the largest confidence-ranked prefix on $D_\tau$ whose
one-sided 95\% Wilson error upper bound is at most $1-0.90$ ($z=1.645$),
retaining ties and freezing the boundary threshold. If none qualifies,
coverage is zero and accuracy undefined. LastFM uses $\alpha=0.25$ above.
The Goodreads development target does not guarantee test accuracy.
\subsection{Confidence-Guided Routing Evaluation}
\label{sec:appendix-routing}
Each run pairs saved 0.6B/4B logits on identical examples and candidates,
fitting small-model temperature on $D_T$ and freezing $\gamma$ on $D_\tau$
(Equations~\ref{eq:fixed-service}, \ref{eq:routing}). Figure~\ref{fig:router}
sweeps common development-retention targets in 5\% steps, averaging
test HR@1 and actual escalation across runs at each target.
The horizontal axis is mean actual test escalation.
The random-escalation reference is
$(1-r_{\mathrm{esc}})a_{0.6}+r_{\mathrm{esc}}a_4$, using mean test escalation
$r_{\mathrm{esc}}$ and each model's mean full-test accuracy.
Estimated sequential latency is
\begin{equation}
  \widehat t_{\mathrm{cascade}}=\bar t_{0.6}+r_{\mathrm{esc}}\bar t_4.
  \label{eq:cascade-cost}
\end{equation}
For the LastFM comparison, the 0.6B tree cost is 38.34 ms
(Figure~\ref{fig:latency}); the 4B tree cost is 139.85 ms
(128 examples, chunk size eight, two warm-up batches), both on H20.
Both component timings include tokenization, input preparation, model
inference, and item selection. The 4B-only reference uses the same 4B tree
path and has cost $\bar t_4$.
The cost model combines mean test escalation with these component means;
it does not measure an online cascade.
The estimate excludes routing overhead; all-escalation incurs both model costs.
\section{Candidate-Interaction Extension (DICE-S)}
\label{sec:appendix-dice-s}
DICE-S adds residual interactions among DICE-I's candidate representations,
using permutation-equivariant set modeling~\cite{zaheer2017deepsets,lee2018settransformer}.
\paragraph{Interaction module.}
For each candidate, we first project its representation into an interaction space:
\begin{equation}
  g_i
  =
  W_g[\,\operatorname{LN}(r_i);\log K\,]+b_g,
  \label{eq:set-projection}
\end{equation}
where $[\,\cdot\,;\,\cdot\,]$ denotes concatenation.
The scalar $\log K$ provides the candidate-set size and is shared by all candidates in the same example.
With $G=[g_1,\ldots,g_K]^\top$, DICE-S applies multi-head self-attention
without candidate-position embeddings:
\begin{equation}
  \tilde g_i
  =
  \operatorname{MHA}(g_i,G,G;M),
  \label{eq:set-attention}
\end{equation}
where $M$ masks padded candidates.
The interaction-aware logit is
\begin{equation}
  z_i^{\mathrm S}
  =
  z_i^{\mathrm I}
  +
  w_S^\top\tanh(g_i+\tilde g_i)+b_S.
  \label{eq:dice-s-logit}
\end{equation}
The residual head is initialized to zero, so DICE-S starts from the DICE-I scores.
DICE-S uses the softmax and choice loss in
Eqs.~\eqref{eq:choice-probability}--\eqref{eq:choice-loss}, updating LoRA,
the independent head, and the interaction module while freezing the backbone.
It shares the single-forward inference layout of Section~\ref{sec:inference}.
\paragraph{Candidate-order consistency.}
Shared projections, attention without candidate-position embeddings,
and masks that follow candidates under reordering preserve permutation
equivariance:
\begin{equation}
  z_i^{\mathrm S}(H,C^\pi)=z_{\pi(i)}^{\mathrm S}(H,C).
  \label{eq:dice-s-permutation}
\end{equation}
Softmax preserves this relation. With DICE-I's item-identifier tie rule,
item selection is unchanged; Appendix~\ref{sec:appendix-order} gives the empirical check.
\paragraph{Empirical comparison.}
Table~\ref{tab:main} and Figure~\ref{fig:rq2-candidate-sensitivity} show
that interaction gains vary across datasets and candidate counts, so DICE-S
remains an optional extension.
\section{Representation and Adaptation Controls}
\label{sec:appendix-representation}
We compare DICE-I with two supplementary controls.
\emph{Frozen} trains only a choice head over frozen Qwen3 text representations.
\emph{ID-only} uses GRU4Rec from Table~\ref{tab:main},
which models item-ID interaction histories without item text or a pretrained LLM.
\begin{table}[!htbp]
  \caption{Mean HR@1 for representation and adaptation controls.
  Best results are bold.}
  \label{tab:representation-controls}
  \centering
  \begin{tabular*}{\columnwidth}{@{\extracolsep{\fill}}lrr@{}}
    \toprule
    Variant & Goodreads & LastFM \\
    \midrule
    Frozen encoder + head-only & 0.2243 & 0.2032 \\
    GRU4Rec (ID-only) & 0.3870 & 0.2617 \\
    DICE-I & \textbf{0.5332} & \textbf{0.6160} \\
    \bottomrule
  \end{tabular*}
\end{table}
DICE-I achieves higher HR@1 than both controls on Goodreads and LastFM
(Table~\ref{tab:representation-controls}). These results support the use
of task-adapted text representations over the frozen-feature
and ID-only alternatives. The controls differ in trainable capacity,
architecture, and available information.

\end{document}